\documentstyle[12pt]{article}
\newcounter{romanlistc}
\newcounter{alphlistc}

\newenvironment{romanlist}
        {\setcounter{romanlistc}{0}
         \begin{list}{$($\roman{romanlistc}$)$}
        {\usecounter{romanlistc}
         \setlength{\parsep}{0pt}
         \setlength{\itemsep}{0pt}}}{\end{list}}

\newenvironment{alphlist}
        {\setcounter{alphlistc}{0}
         \begin{list}{$($\alph{alphlistc}$)$}
        {\usecounter{alphlistc}
         \setlength{\parsep}{0pt}
         \setlength{\itemsep}{0pt}}}{\end{list}}

\newcommand{\nc}{\newcommand}
\nc{\bb}{\begin{equation}}
\nc{\ee}{\end{equation}}
\nc{\r}{\rho}
\nc{\be}{\beta}
\nc{\ra}{\rangle}
\nc{\la}{\langle}
\nc{\Ol}{\la O(\be)|}
\nc{\Or}{| O(\be) \ra}
\nc{\Ort}[1]{| O(\be , #1) \ra}
\nc{\tr}{\,{\rm tr}\,}
\renewcommand{\Im}{\,\mbox{\rm Im}\,}

\nc{\brls}{\begin{romanlist}}
\nc{\erls}{\end{romanlist}}
\nc{\bals}{\begin{alphlist}}
\nc{\eals}{\end{alphlist}}
\nc{\res}{\,{\rm res}\,}
\nc{\bbb}{\begin{eqnarray}}
\nc{\eee}{\end{eqnarray}}
\nc{\bfig}{\begin{figure}}
\nc{\efig}{\end{figure}}
\nc{\bpic}{\begin{picture}}
\nc{\epic}{\end{picture}}
\nc{\ct}{\cite}
\title{The Problem of Ground State in Thermo-Field Dynamics.}
\author{A.A.Abrikosov (Jr.)
\thanks{Talk given at 3rd Workshop on Thermal Field Theories, August 1993,
Banff, Canada.}
\\ {\em Institute for Theoretical and Experimental Physics,} \\
B.Cheremushkinskaya str. 25, 117259 Moscow, RUSSIA}

\begin{document}
\maketitle

\begin{abstract}
Analytic continuation of quantum statistical physics from
imaginary to real time is analyzed. Adiabatic vanishing of interactions
at real time infinities gives origin to singularities at complex
times. This undermines the hypothesis of decoupling of interactions at
$t \rightarrow \infty$. Hence an interacting thermal vacuum is a necessary
component of the exact real-time formalism. Consequences for TFD are
discussed.
\end{abstract}

\section{Introduction}  

Usually quantum statistical physics is formulated in terms of density
matrix (DM). In equilibrium it is given by Gibbs formula \ct{LL}
\bb
\rho = \sum _{n=0}^{\infty} |n \rangle \exp{-\beta E_n} \langle n|,
\ee
$\beta$ being inverse temperature and $E_n$ standing for the energy of the
$n$-th level. The trace of $\rho$ defines thermodynamic functions and kinetic
properties are governed by the evolution of DM \ct{LP}.

The idea of thermo-field dynamics (TFD) is to consider instead a quantum
field theory formulated in a special thermal vacuum (TV) \ct{UMT,VW} $\Or$.
Expectation values should coincide with the averages calculated by means of
DM.
\bb
\hat{A}=
\Ol \hat{A} \Or = \tr \r \hat{A}.
\ee
Kinetic properties are defined by
time dependent correlation functions in TV.
\bb
\la \hat{A}(t)\hat{B}(0)
\ra = \Ol \hat{A} \exp (-i \hat{H} t) \hat{B} \Or.
\ee
It's well known that
the number of degrees of freedom is doubled in TFD: every  physical field
$\phi$ enters with the corresponding "ghost" $\tilde{\phi}$.
Physical density matrix is obtained after the convolution of $\Or \Ol$ with
respect to the tilde-fields.
\bb
\r =\tilde{\tr}\Or \Ol
\ee

The obvious advantage is that TFD looks like the ordinary field theory. Real
time formalism (RTF) immediately extends the scope to kinetic problems.  The
less transparent benefit is cancellation of higher powers of
$\delta$-functions \ct{NS,VW}. Those appear when one takes into account beads
of self-energy corrections to propagators of real particles in the heat
bath.  Fortunately the contribution of ghosts cancels poorly defined terms.

The structure of TFD and the origin of ghosts are much more transparent  in
the time-path (TP) method \ct{NS}. It claims that TFD and familiar Matsubara
technique are linked by an analytic continuation. The subject of the talk
are problems met by this method in interacting theories. The procedure looks
safe at the level of free fields. However in presence of perturbations there
are singularities in complex time plane restricting deformations of
TP \ct{AAA}. Physically this reflects the difference between interacting and
noninteracting thermal vacua.

\section{Time-Path Method}

We shall sketch the analytic continuation relating TFD to Matsubara
imaginary-time formalism \mbox{\ct{NS,VW}}. The free energy of a quantum
system is
\bb F= - {1 \over \be} \log \tr \exp{-\be \hat{H}}
\ee

In Matsubara approach it is treated as a quantum field generating
functional \ct{AGD,RF} for the evolution during finite imaginary time $-i
\be$.  Bose-fields are $\be$ periodic and Fermi-fields are antiperiodic in
this coordinate. The time interval of evolution is portrayed by the contour
$C^M$ in complex time plane (see Fig. 2.).

It proves that $F$ is invariant with respect to deformations preserving
monotonous decrease of $\Im t$ along the contour \ct{Mls}. Consider the
deformation shown in Fig. 1.
\bfig
\setlength{\unitlength}{1mm}
\bpic(90,50)(-30,0)
\put(5,5){\line(1,0){80}}
\put(5,20){\line(1,0){80}}
\put(5,35){\line(1,0){80}}
\put(45,5){\vector(0,1){40}}
\multiput(45,5)(0,15){3}{\circle*{.5}}
\put(25,35){\vector(1,0){0}}
\put(65,35){\vector(1,0){0}}
\put(25,20){\vector(-1,0){0}}
\put(65,20){\vector(-1,0){0}}
\put(80,25){\vector(0,-1){0}}
\put(10,10){\vector(0,-1){0}}
\put(49,9){\makebox(0,0){$-i \be$}}
\put(49,24){\makebox(0,0){$-i \frac{\be}{2}$}}
\put(49,39){\makebox(0,0){$0$}}
\put(35,40){\makebox(0,0){$t$}}
\put(10,40){\makebox(0,0){$-T$}}
\put(80,40){\makebox(0,0){$T$}}
\put(35,40){\circle{4}}
\thicklines
\put(10,20){\line(1,0){70}}
\put(10,35){\line(1,0){70}}
\put(10,5){\line(0,1){15}}
\put(80,20){\line(0,1){15}}
\put(15,15){\makebox(30,5){"ghost axis"}}
\put(10,5){\circle*{1}}
\put(10,35){\circle*{1}}
\epic
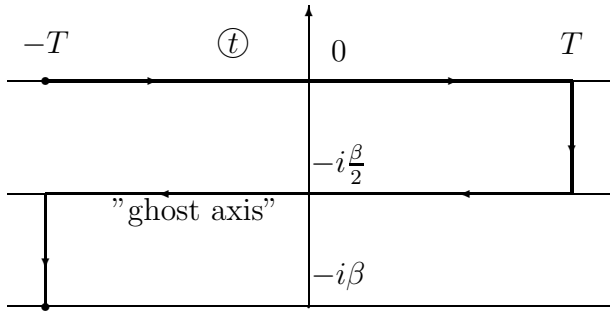
\caption{The contour $C_{TFD}$ used to derive real-time formalism. Physical
fields live on the real time axis and tilde-fields are  defined on the
"ghost axis" $\Im t= - \be /2$.}
\efig
It turns out that in the limit $T \rightarrow \infty$ for free
fields one obtains the formulae of TFD. Physical fields are defined on the
real time axis and "tilde"-fields live on the "ghost axis" $\Im t= - \be
/2$.

The usual assumption is that the vertical pieces decouple as
$T \rightarrow \infty$. That infers that they result only into a
multiplicative renormalization of the statistical sum and do not contribute
significantly to  thermodynamic functions.

\section{Adiabatic perturbations and complex time singularities}

Usually one implies in QFT that interaction is absent at $t= \pm \infty$ and
turns on adiabatically in physical domain. Thus asymptotic states are free.
They form a complete set used as a basis in calculations.

Decoupling of vertical sections of the TFD-path also suggests that the
asymptotic states are the same as without perturbation. So the latter might
be turned off adiabatically at real time infinities. However in thermal
theories these adiabatic changes give rise to singularities of the
perturbation at complex times \ct{AAA}. The latter restrict deformation of the
time-path making noninteracting area inaccessible. This indicates that
interaction modifies the ground state and thermal vacuum may differ
substantially from the set of free thermal quanta.

To prove the existence of singularities let's recall the general properties
of perturbation. It's natural to believe that:
\bals
\item  Perturbation $V$ is real on the real time axis and because of
analyticity $V(\bar{t})= \bar{V}(t)$. (Bar stands for complex conjugation.)
\item In physical region $V(|t| \ll T) \approx const $ and $V(t=\pm
\infty)=0 $.
\item  Perturbation $V(t)$ is periodic in imaginary time with the
period $i \be $.
\eals
The last requirement means that $V(t)$ has the same temperature as the
nonperturbed system. Otherwise we encounter a nonequilibrium situation and
heating processes occur.

The TPM is applicable if $V(t)$ is analytic throughout the strip $- \be <\Im
t<0$.  Meanwhile the requirements (b) and (c) are fulfilled simultaneously
only for functions with singularities in this region. One can prove this in
the following way.
\brls
\item Periodicity makes the domain of definition of $V$
topologically equivalent to an infinite cylinder.
\item The condition (b) permits to add points
$t=\pm \infty $ converting the cylinder to a sphere.
\item The number of
zeros of a function analytic on a sphere is equal to that of poles. Hence
$V$ has in the strip {\bf at least two simple poles}. (The case of a second
order pole contradicts $V \approx const $ in physical region.)
\footnote{I shall call these poles Hoo-Doos after the ghosts watching the
horizons of Banff on foggy days.}
\erls

Hereon we shall discuss this simplest case.

It is easy to show that the singularities are symmetric with respect to the
ghost axis $\Im t=-\be /2$.
\brls
\item According to (a) there should be a pole 2 at $\Im t =
\frac{\be}{2}+\Delta$ if there is a pole 1 at $\Im t =
-\frac{\be}{2}-\Delta$.
\item Periodicity immediately requires the
existence of a pole 3 at $\Im t = -\frac{\be}{2}+\Delta$.
\item The poles 1 and 3 are symmetric with respect to the ghost axis.
\erls
Proceeding in an analogous way one can prove that $V(t)$ must be real on the
ghost axis.

We can find residues of $V$ at the poles (or the integral over the border of
the analytic domain for more elaborate singularities, Fig.4). To do that we
shall integrate $V$ along the contours $C_1$, $C_2$ shown in Fig.2.
\bfig
\setlength{\unitlength}{1mm}
\bpic(90,50)(-30,0)
\put(5,5){\line(1,0){80}}
\put(5,35){\line(1,0){80}}
\put(45,5){\vector(0,1){40}}
\multiput(45,5)(0,30){2}{\circle*{1}}
\put(20,35){\vector(-1,0){0}}
\put(70,35){\vector(-1,0){0}}
\put(35,22){\vector(0,1){0}}
\put(45,20){\vector(0,-1){0}}
\put(55,20){\vector(0,-1){0}}
\put(20,5){\vector(1,0){0}}
\put(70,5){\vector(1,0){0}}
\put(49,9){\makebox(0,0){$-i \be$}}
\put(49,39){\makebox(0,0){$0$}}
\put(35,40){\makebox(0,0){$t$}}
\put(35,40){\circle{4}}
\thicklines
\put(5,5){\line(1,0){30}}
\put(55,5){\line(1,0){30}}
\put(5,35){\line(1,0){30}}
\put(55,35){\line(1,0){30}}
\put(35,5){\line(0,1){30}}
\put(45,5){\line(0,1){30}}
\put(55,5){\line(0,1){30}}
\multiput(20,20)(50,0){2}{\circle*{2}}
\put(20,38){\makebox(0,0){$C_1^+$}}
\put(70,38){\makebox(0,0){$C_2^+$}}
\put(20,8){\makebox(0,0){$C_1^-$}}
\put(70,8){\makebox(0,0){$C_2^-$}}
\put(31,20){\makebox(0,0){$C_1^M$}}
\put(49,20){\makebox(0,0){$C^M$}}
\put(59,20){\makebox(0,0){$C_2^M$}}
\put(20,23){\makebox(0,0){$HD_1$}}
\put(70,23){\makebox(0,0){$HD_2$}}
\multiput(5,5)(0,4){8}{\line(0,1){2}}
\multiput(85,5)(0,4){8}{\line(0,1){2}}
\put(9,20){\makebox(0,0){$C_1^{\infty}$}}
\put(81,20){\makebox(0,0){$C_2^{\infty}$}}
\put(85,22){\vector(0,1){0}}
\put(5,20){\vector(0,-1){0}}
\epic
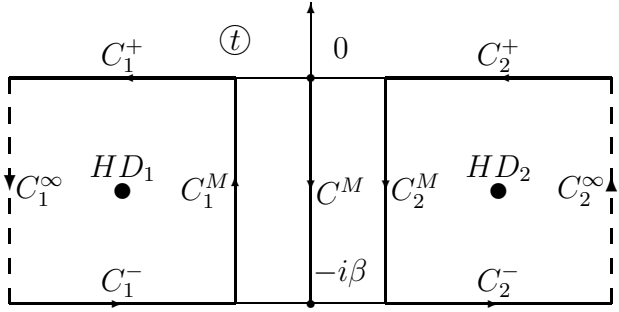
\caption{Matsubara contour $C^{M}$ and contours $C_{1,2}$ used to calculate
the residues at the poles $HD_{1,2}$.}
\efig
The integral can be split into
a sum of the four pieces.
\bb
\oint \limits_C = \int
\limits_{C^M}+\int \limits_{C^-}+ \int \limits_{C^+} + \int
\limits_{C^{\infty}}
\ee
The last is zero because of (b) and periodicity is a
reason for cancellation of the 2-nd with the 3-rd. Thus
\bb
\res _{1(2)}
V(t)= (-) \frac{1}{2 \pi i} \int \limits_{C_{M}} V(t)\,dt = (-)(-\frac{\be
V}{2 \pi})
\ee
The result is absolutely natural. (Note that the nonzero
residues are an independent proof of the existence of singularities.)

Finally let us give an example of a function possessing all the listed
properties (a)--(c).
\bb
V(t) = V_0 \frac{1+ \cosh \frac{2 \pi
t_{HD}}{\be}} {\cosh \frac{2 \pi t}{\be}+ \cosh \frac{2 \pi t_{HD}}{\be}}
\ee
It is $i \be$ periodic and has poles at the points $\pm t_{HD}+ i \be (n-
\frac{1}{2})$. It vanishes at real time infinities and $V(|t| \ll t_{HD})
\approx V_0$. The residues are $\frac{\be}{2 \pi}V(0)+ O(e^{-\frac{4 \pi
t_{HD}}{\be}})$.

\section{The problem of interacting thermal vacuum}

The complex time singularities restrict deformations of time paths, Fig.3.
\bfig
\setlength{\unitlength}{1mm}
\bpic(90,50)(-30,0)
\put(5,5){\line(1,0){80}}
\put(5,20){\line(1,0){80}}
\put(5,35){\line(1,0){80}}
\put(45,5){\vector(0,1){40}}
\multiput(45,5)(0,15){3}{\circle*{.5}}
\put(35,35){\vector(1,0){0}}
\put(60,35){\vector(1,0){0}}
\put(30,20){\vector(-1,0){0}}
\put(55,20){\vector(-1,0){0}}
\put(70,25){\vector(0,-1){0}}
\put(20,10){\vector(0,-1){0}}
\put(49,9){\makebox(0,0){$-i \be$}}
\put(49,24){\makebox(0,0){$-i \frac{\be}{2}$}}
\put(49,39){\makebox(0,0){$0$}}
\put(35,40){\makebox(0,0){$t$}}
\put(35,40){\circle{4}}
\thicklines
\put(20,20){\line(1,0){50}}
\put(20,35){\line(1,0){50}}
\put(20,5){\line(0,1){15}}
\put(70,20){\line(0,1){15}}
\put(20,35){\circle*{1}}
\put(20,5){\circle*{1}}
\put(10,20){\circle*{2}}
\put(80,20){\circle*{2}}
\put(10,23){\makebox(0,0){$HD_1$}}
\put(80,23){\makebox(0,0){$HD_2$}}
\put(20,40){\makebox(0,0){$-T$}}
\put(70,40){\makebox(0,0){$T$}}
\epic
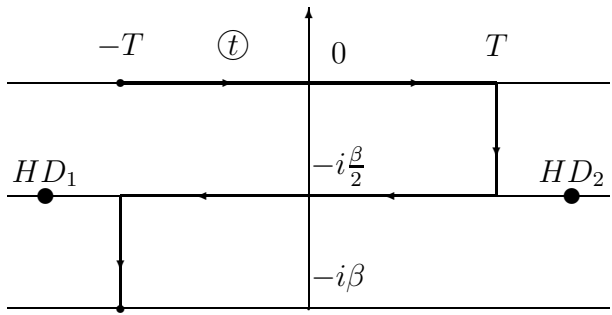
\caption{Singularities of perturbation restrict deformations of the time
path so that $T \ll t_{HD}$.}
\efig

In fact this evidences that thermal theories are selfconsistent only if the
vacuum state is interacting. As long as the vertical sections of the
RTF-contour lay in the physical domain where $V \neq 0$ one deals indeed
with an analytic continuation of the imaginary time Matsubara approach.
However neglecting the contribution of the vertical pieces is not an
analytic procedure.

Strictly speaking that means that interacting TFD is not an analytic
continuation of imaginary time methods. A question arises what are the cases
when this beautiful theory gives good approximations and when the
discrepancy with conservative approaches becomes essential.

The complete analytic continuation begins with the solution of quantum
statistical (Matsubara) problem. The difference comes forth from the
dependence of thermal vacuum on perturbation.
\bbb
\Or_V \propto \sum_n \exp{-\frac{\be
E_{n}}{2}}|n \ra _V |\tilde{n} \ra _V & \neq & \Or_0 \propto \sum_n
\exp{-\frac{\be E^0_{n}}{2}}|n \ra _0 |\tilde{n} \ra _0
\eee
One can distinguish two effects caused by the interaction. The first is
shifting energy levels and the second is the related change of occupation
numbers $\exp - \be E_n^0 \rightarrow \exp - \be E_n$. Dynamical effects of
the level shifting are taken into account by the horizontal parts of the
path $C_{TFD}$, Fig.1, whereas the vertical sections are responsible for the
corrections to occupation numbers.

In TFD one makes use of the noninteracting thermal vacuum obtained from the
zero temperature one by means of Bogolyubov rotation.
\bb
\Or_{TFD} =\Or_0
= B | O(\infty) \ra
\ee
This means neglecting the second effect
\footnote{Note that Green functions $G(t_1 , t_2 )$ are defined as averages
in noninteracting vacuum. Hence switching off actually takes place at {\em
finite} times $t_1 $, $t_2 $. This makes one more difference from $T
\rightarrow \infty $ limit considered for a free theory.}.

As a result the calculated values of the
free energy $F_{TFD}$ and of the entropy $S_{TFD}$ differ from the true $F$
and $S$.  According to LeCh\^{a}telieu-Brown principle \ct{LL,RF}
\bbb
F \leq F_{TFD} = F_0 + \la V \ra_0; & & S_{TFD} \leq S
\eee

Analysis of the second inequality makes up the matters with
the common belief that adiabatic perturbations do not spoil the state of
the system.  The point is that a realistic system would respond to
variations of the perturbation by relaxation processes. The latter break
analyticity and the corresponding complex time plane is shown in Fig.4.
\bfig
\setlength{\unitlength}{1mm}
\bpic(90,50)(-30,0)
\put(5,5){\line(1,0){80}}
\put(5,35){\line(1,0){80}}
\put(45,25){\vector(0,1){20}}
\put(45,5){\line(0,1){10}}
\put(10,5){\line(0,1){30}}
\put(30,5){\line(0,1){30}}
\put(60,5){\line(0,1){30}}
\put(80,5){\line(0,1){30}}
\multiput(10,5)(5,0){3}{\line(1,1){10}}
\multiput(10,25)(5,0){3}{\line(1,1){10}}
\multiput(60,25)(5,0){3}{\line(1,1){10}}
\multiput(60,5)(5,0){3}{\line(1,1){10}}
\put(10,10){\line(1,1){5}}
\put(25,5){\line(1,1){5}}
\put(10,30){\line(1,1){5}}
\put(25,25){\line(1,1){5}}
\put(60,10){\line(1,1){5}}
\put(75,5){\line(1,1){5}}
\put(60,30){\line(1,1){5}}
\put(75,25){\line(1,1){5}}
\multiput(45,5)(0,15){3}{\circle*{.5}}
\put(49,9){\makebox(0,0){$-i \be$}}
\put(49,39){\makebox(0,0){$0$}}
\put(35,40){\makebox(0,0){$t$}}
\put(35,40){\circle{4}}
\put(10,15){\makebox(20,10){\shortstack{relaxation\\zone}}}
\put(60,15){\makebox(20,10){\shortstack{relaxation\\zone}}}
\put(45,20){\makebox(0,0){\shortstack{interaction\\zone}}}
\put(20,38){\makebox(0,0){$\Delta S > 0 $}}
\put(70,38){\makebox(0,0){$\Delta S > 0 $}}
\epic
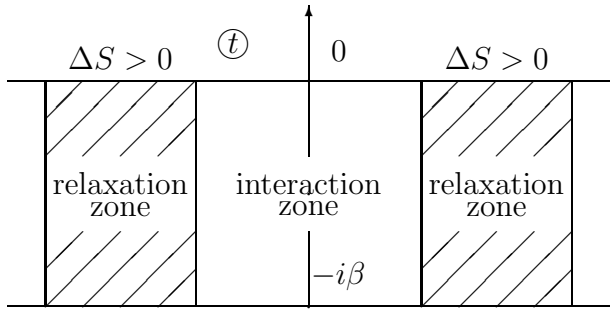
\caption{In realistic systems analyticity is broken by relaxation
processes.}
\efig
Entropy production in relaxation processes leads to the
inequality (10). Note that the poles we are discussing are just the
fingerprints of the nonanalytic relaxation zones.

Analysis of thermodynamic potential \ct{LL,AGD} $\Omega=-pv$ ($p$ and $v$
being pressure and volume) indicates that values of pressure in interacting
and noninteracting vacua differ by \ct{AAA}

\bb
\Delta p=-\frac{\Delta \Omega}{v}= \frac{1}{\be v} \log \tr (- \be \hat{V}).
\ee

\section{Discussion and examples}

Now we shall briefly discuss some cases where sensitivity to the interaction
in the initial state can be important. Somehow all listed here happen to
deal with some aspects of the ground state symmetry.

The first example are theories with spontaneously broken symmetry ({\em e.g.}
$\lambda \varphi^4$-theories). It is well known that quantum corrections
make effective potential temperature dependent \ct{AL}. At high temperatures
symmetry restoration takes place. Thus the ground state ({\em i.e.} thermal
vacuum) depends both on temperature and interaction. The naive approach is
unsuitable in this case.

Interaction can form new states in the energy spectrum. (The arising
difficulties are not specific to TFD alone.) If it is the case the
noninteracting basis is incomplete and the theory can not be formulated in
terms of the free fields. The classic example is that of
superconductivity where the existence of condensate is an assumption
necessary for a successful treatment.

New states with high energy are not so important and the most prominent
contributions come from zero modes, {\em i.e.} states $| \psi_0 \ra$ with
zero energy: $(\hat{H} +\hat{V})| \psi_0 \ra =0$. These states violate
conditions of Riemann-Lebesgue lemma which is crucial in the proof of
decoupling \ct{VW}.

The more general are cases of so called quantum algebras \ct{QA}. (These
should be not mixed with those studied in connection with string and
conformal field theories.) Here the interacting thermal vacuum belongs to
nontrivial representation of the corresponding group whereas the free one by
definition is a singlet.  The transformation properties impose selection
rules on expectation values.  This sort of situation was found to take place
in Anderson model \ct{JH} where $\Or=\otimes_i \Or_i $ and operators
from different sectors enter calculations accompanied by different sets of
closed vacuum graphs.

Finally we shall discuss a simple case where the assumption about initial
state immediately affects the calculation. Imagine a system of
noninteracting $\frac{1}{2}$-spins in a magnetic field $h$ which is treated
as a perturbation. (This generalizes to arbitrary degenerate levels which
are split by some interaction.) It happens that pure quantum mechanical
evolution which begins from noninteracting ({\em i.e.} degenerate) state is
unable to provide a nonzero magnetization in this system.

The hamiltonian $\hat{H}$  and the perturbation $\hat{V}$ are
\bbb
\hat{H_0}=0; & & \hat{V_0}=-\frac{1}{2} \sigma_z h;
\eee
The degenerate {\em in}-state is written as follows (we shall proceed using
both TFD and DM formalisms).
\bbb
\Ort{0} =\frac{1}{\sqrt{2}}[|\uparrow \tilde {\uparrow} \ra + |\downarrow
\tilde {\downarrow} \ra ]; & &
\r(0)=\frac{1}{2} \left(
\begin{array}{cc} 1 & 0 \\ 0 & 1
\end{array}
\right)
\eee
Their  evolutions are governed by the full hamiltonians
\bbb
\hat{H}_{TFD} = \hat{V} -\hat{ \tilde {V}}; & & \hat{H}_{DM} =\hat{V}.
\eee
The laws of evolution are:
\bbb
\Ort{t} = \exp (-i \hat{H}_{TFD} t) \Ort{0}; & &
\r (t) = \exp (-i \hat{H}_{DM}t) \r (0)\exp (i \hat{H}_{DM}t)
\eee
Without much effort one sees that $\Ort{t}$ and $\r (t)$ do not change and
the states remain degenerate despite the interaction. Hence there is no
magnetization in this approach which certainly is wrong.

Meanwhile taking the perturbation into account from the very beginning
gives:
\bbb
\Or_V = \frac{| e^{\frac{\be h}{4}}\uparrow \tilde {\uparrow} \ra +
e^{-\frac{\be h}{4}}|\downarrow \tilde {\downarrow} \ra }{\sqrt
{2 \cosh \be h}}; & &
\r _V = \frac{1}{2 \cosh \frac{\be h}{2}}
\left(
\begin{array}{cc} e^{\frac{\be h}{2}} & 0 \\ 0 & e^{-\frac{\be h}{2}}
\end{array}
\right)
\eee
Which leads to the known value
\bb
M_z = \frac{1}{2} \, \mbox{\rm tanh} \, \frac{\be h}{2}
\ee
The example demonstrates that some physical quantities are especially
sensitive to particular details of the structure of thermal vacuum. In our
case magnetization was the one feeling the degeneracy of states. Quantum
mechanics disregards relaxation and systems have a memory of the initial
state. This is not the case for statistical physics. Generally some special
improvements of thermal vacuum should be made by hand in order to obtain
the correct values of sensitive quantities.

\section{Summary}  

We have shown that adiabatically changing perturbations can not be safely
incorporated in thermal theory framework. For interacting theory the real
time formalism  has been shown not to be an analytic continuation of the
imaginary time technique. That means that TFD is not a new guise of the
Matsubara approach. The use of noninteracting thermal vacuum is an
approximation which needs justification in particular cases. The reason for
that is not simply striving for mathematical rigorousness but sensitivity of
certain physical quantities to the details of the structure of the ground
state.

The tool for analyzing a thermal vacuum is the interacting imaginary time
formalism. With the help of it one can find out if the nonperturbed TFD
vacuum is good enough for a specific problem and decide what should be the
necessary {\em a priori} improvements. With the corrected vacuum state all
the power and beauty of TFD are welcome and the circumspect ones may no
longer be afraid of Hoo-Doos on the chosen trail.

\section{Acknowledgements}  
My participation in the Banff workshop was sponsored by the International
Science Foundation. I'd like to thank the my colleagues and staff of the
Institute for Theoretical Physics where I had a pleasure to prepare the
final version. That was supported in part by National Science Foundation
under Grant No.~PHY~89-04035.


\begin{thebibliography}{99}
\bibitem{LL}
L.~D.~Landau and E.~M.~Lifshitz, {\it Statistical Physics} (Pergamon Press,
Oxford, 1980).
\bibitem{LP}
E.~M.~Lifshitz,  L.~P.~Pitayevsky, {\it Physical Kinetics} (Pergamon Press,
Oxford, 1981).
\bibitem{UMT}
H.~Umezawa, H.~Matsumoto, M.~Tachiki, {\it Thermo-Field Dynamics and
Condensed States} (North Holland, Amsterdam, 1982).
\bibitem{VW}
N.~P.~Landsman, Ch.~G.~van~Weert, {\it Phys. Rep. C,} {\bf 145}, (1987) 141.
\bibitem{NS}
A.~J.~Niemi, G.~W.~Semenoff, {\it Ann. Phys.,} {\bf 152} (1984) 105; {\it
Nucl.  Phys.,} {\bf B230} (1984) 181.
\bibitem{AAA}
A.~A.~Abrikosov Jr., {\it Mod. Phys. Lett. A}, {\bf 5} (1990) 2183.
\bibitem{AGD} A.~A.~Abrikosov, L.~P.~Gorkov, I.~E.~Dzyaloshinski,
{\em Methods of Quantum Field Theory in Statistical Physics} (Prentice-Hall,
1963)
\bibitem{RF}
R.~P.~Feynman, {\it Statistical mechanics} (W.~A.~Benjamin, Inc., Massachusets,
1972).
\bibitem{Mls}
R.~Mills, {\it Propagators for many particle systems,} (Gordon \& Breach,
New York, 1962).
\bibitem{AL}
A.~D.~Linde, {\it Rep. Progr. Phys.,} {\bf 42} (1979) 389.
\bibitem{QA}
H.~Matsumoto, H.~Umezawa, {\it Phys. Lett.,} {\bf 103A} (1984) 405.
\bibitem{JH}
J.~Hebron, private communication.
\end{thebibliography}
\end{document}